            \documentstyle[epsf,12pt]{article}
\begin{document}\thispagestyle{empty}\begin{flushright}
UTAS--PHYS--96--44\\OUT--4102--64\\MZ--TH/96--25\\
revised November 14, 1996\\change to Eq (7)\\change to item 1.~of conclusions
                         \\update to refs.~[6,9,10,14]
            \end{flushright} \vspace*{2mm} \begin{center} {\large\bf
Association of multiple zeta values with positive knots\\
via Feynman diagrams up to 9 loops$^{*)}$}\vglue 6mm{\large{\bf
D.~J.~Broadhurst}$^{1)}$ and {\bf
D.~Kreimer}$^{2)}$}\vglue 4mm
Department of Physics, University of Tasmania,\\
GPO Box 252C, Hobart, Tasmania 7001, Australia\end{center} \vfill
{\bf Abstract}
It is found that the number, $M_n$, of irreducible multiple zeta values (MZVs)
of weight $n$, is generated by $1-x^2-x^3=\prod_n (1-x^n)^{M_n}$.
For $9\ge n\ge3$, $M_n$ enumerates positive knots with $n$ crossings.
Positive knots to which field theory assigns knot-numbers that are not MZVs
first appear at 10 crossings. We identify all the positive knots, up to
15 crossings, that are in correspondence with irreducible MZVs, by virtue of
the connection between knots and numbers realized by Feynman diagrams
with up to 9 loops.
                                                    \vfill
                                                    \footnoterule\noindent
$^*$) Work supported in part by grant CHRX--CT94--0579, from HUCAM.\\
$^1$) D.Broadhurst@open.ac.uk;
      on leave of absence from the Open University, UK.\\
$^2$) kreimer@dipmza.physik.uni-mainz.de;
      on leave of absence from Mainz University, FRG.
\newpage
\setcounter{page}{1}
\newcommand{\df}[2]{\mbox{$\frac{#1}{#2}$}}
\newcommand{\Eq}[1]{~(\ref{#1})}
\newcommand{\Eqq}[2]{~(\ref{#1},\ref{#2})}
\newcommand{\Eqqq}[3]{~(\ref{#1},\ref{#2},\ref{#3})}
\newcommand{\la}{\lambda}

\subsection*{1. Introduction}

The connection of positive knots with transcendental numbers, via the
counterterms of quantum field theory, proposed in~\cite{DK1} and
developed in~\cite {DK2}, and has been vigorously tested against
previous~\cite{GPX,DJB} and new~\cite{BKP} calculations,
entailing knots with up to 11 crossings, related by counterterms with
up to 7 loops to numbers that are irreducible
multiple zeta values (MZVs)~\cite{DZ,LM}.
Cancellations of transcendentals in gauge
theories have been illuminated by knot theory~\cite{BDK}. All-order results,
from large-$N$ analyses~\cite{BGK} and Dyson-Schwinger methods~\cite{DKT},
have further strengthened the connection of knot theory and number theory,
via field theory. A striking feature of this connection is that the
first irreducible MZV of depth 2 occurs at weight 8~\cite{DJB,BBG}, in
accord with the appearance of the first positive 3-braid knot at crossing
number 8. Likewise the first irreducible MZV of depth 3 occurs at weight
11~\cite{BG}, matching the appearance of the first positive 4-braid
at 11 crossings, obtained from skeining link diagrams that encode momentum
flow in 7-loop counterterms~\cite{BKP}. Moreover, the
investigations in~\cite{BGK} led to a new discovery at weight 12,
where it was found that the reduction of MZVs first entails alternating
Euler sums. The elucidation of this phenomenon
resulted in an enumeration~\cite{EUL} of irreducible Euler sums
and prompted intensive searches for evaluations of sums of
arbitrary depth~\cite{BBB}. A review of all these developments is
in preparation~\cite{DK}.

This paper pursues the connection to 8 and 9 loops, entailing knots
with up to 15 crossings.
In Section~2, we enumerate irreducible MZVs by weight.
Section~3 reports calculations of Feynman diagrams that yield
transcendental knot-numbers entailing MZVs up to weight 15.
In Section~4 we enumerate positive knots, up to 15 crossings,
and give the braid words and HOMFLY polynomials~\cite{VJ} for
all knots associated with irreducible MZVs of weight $n<17$.
Section~5 gives our conclusions.

\subsection*{2. Multiple zeta values}

We define $k$-fold Euler sums~\cite{BBG,BG} as in~\cite{EUL,BBB},
allowing for alternations of signs in
\begin{equation}
\zeta(s_1,\ldots,s_k;\sigma_1,\ldots,\sigma_k)=\sum_{n_j>n_{j+1}>0}
\quad\prod_{j=1}^{k}\frac{\sigma_j^{n_j}}{n_j^{s_j}}\,,\label{form}
\end{equation}
where $\sigma_j=\pm1$, and the exponents $s_j$ are positive integers,
with $\sum_j s_j$ referred to as the weight (or level) and $k$ as the depth.
We combine the strings of exponents and signs
into a single string, with $s_j$ in the $j$th position when $\sigma_j=+1$,
and $\overline{s}_j$ in the $j$th position when $\sigma_j=-1$.
Referring to non-alternating sums as MZVs~\cite{DZ},
we denote the numbers of irreducible Euler sums
and MZVs by $E_n$ and $M_n$, at weight $n$, and find that
\begin{equation}
1-x  -x^2=\prod_{n>0}(1-x^n)^{E_n}\,;\quad
1-x^2-x^3=\prod_{n>0}(1-x^n)^{M_n}\,,\label{EM}
\end{equation}
whose solutions, developed in Table~1, are given in closed form by
\begin{eqnarray}
E_n=\frac{1}{n}\sum_{d|n}\mu(n/d)L_d\,;
&&L_n=L_{n-1}+L_{n-2}\,;\quad L_1=1\,;\quad L_2=3\,,\label{Es}\\
M_n=\frac{1}{n}\sum_{d|n}\mu(n/d)P_d\,;
&&P_n=P_{n-2}+P_{n-3}\,;\quad P_1=0\,;\quad P_2=2;\quad P_3=3\,,\label{Ms}
\end{eqnarray}
where $\mu$ is the M\"obius function, $L_n$ is a Lucas number~\cite{EUL} and
$P_n$ is a Perrin number~\cite{AS}.

\noindent{\bf Table~1}:
The integer sequences\Eqqq{Es}{Ms}{Kn} for $n\leq20$.
\[\begin{array}{|r|rrrrrrrrrrrrrrrrrrrr|}\hline
n&1&2&3&4&5&6&7&8&9&10&11&12&13&14&15&16&17&18&19&20\\\hline
E_n&1&1&1&1&2&2&4&5&8&11&18&25&40&58&90&135&210&316&492&750\\
M_n&0&1&1&0&1&0&1&1&1&1&2&2&3&3&4&5&7&8&11&13\\
K_n&0&0&1&0&1&1&1&2&2&3&4&5&7&9&12&16&21&28&37&49\\\hline
\end{array}\]

In~\cite{EUL}, $E_n=\sum_k E_{n,k}$ was apportioned, according to
the minimum depth $k$ at which irreducibles of weight $n$ occur.
Similarly, we have apportioned $M_n=\sum_k M_{n,k}$.
The results are elements of Euler's triangle~\cite{EUL}
\begin{equation}
T(a,b)=\frac{1}{a+b}\sum_{d|a,b}\mu(d)\,P(a/d,b/d)\,,
\label{ET}
\end{equation}
which is a M\"obius transform of Pascal's triangle, $P(a,b)={a+b\choose a}
=P(b,a)$.
We find that
\begin{equation}
E_{n,k}=T(\df{n-k}{2},k)\,;\quad
M_{n,k}=T(\df{n-3k}{2},k)\,,
\label{EMb}
\end{equation}
for $n>2$ and $n+k$ even.
There is a remarkable feature of the result for $M_{n,k}$: it gives
the number of irreducible Euler sums of weight $n$ and depth $k$ that occur
in the reduction of MZVs, which is {\em not\/} the same as the
number of irreducible MZVs of this weight and depth. It was shown
in~\cite{BGK,EUL} that alternating multiple sums occur in the reduction of
non-alternating multiple sums. For example, $\zeta(4,4,2,2)$
cannot be reduced to MZVs of lesser depth, but it can~\cite{EUL} be reduced
to the alternating Euler sum $\zeta(\overline9,\overline3)$.
Subsequently we found
that an analogous ``pushdown'' occurs at weight 15, where depth-5 MZVs,
such as $\zeta(6,3,2,2,2)$, cannot be reduced to MZVs of lesser depth,
yet can be reduced to alternating Euler sums, with
$\zeta(9,\overline3,\overline3)-\frac{3}{14}\zeta(7,\overline5,\overline3)$
serving as the corresponding depth-3 irreducible. We conjecture
that the number, $D_{n,k}$, of MZVs of weight $n$ and depth $k$
that are not reducible to MZVs of lesser depth
is generated by
\begin{equation}
1-\frac{x^3 y}{1-x^2}+\frac{x^{12}y^2(1-y^2)}{(1-x^4)(1-x^6)}
=\prod_{n\ge3} \prod_{k\ge1} (1-x^n y^k)^{D_{n,k}},\label{Pd} 
\end{equation}
which agrees with~\cite{BBG,BG} for $k<4$ and all weights $n$,
and with available data on MZVs, obtained from binary
reshuffles~\cite{LM} at weights $n\leq20$ for $k=4$, and $n\leq18$
for $k>4$. Further tests of\Eq{Pd} require very large scale
computations, which are in progress, with encouraging results.
However, the work reported here does not rely on this conjecture;
the values of $\{M_n\mid n\le15\}$ in Table~1 are sufficient for present
purposes, and these are amply verified by exhaustive use of
the integer-relation search routine MPPSLQ~\cite{DHB}.

Finally, in this section, we remark on the simplicity of the prediction
of\Eq{Ms} for the dimensionality, $K_n$, of the search space
for counterterms that evaluate to MZVs of weight $n$.
Since $\pi^2$, with weight $n=2$, does
not occur in such counterterms, it follows that they
must be expressible in terms of transcendentals that are enumerated by
$\{M_n\mid n\geq3\}$, and products of such knot-numbers~\cite{DK1,BGK,EUL}.
Thus $K_n$ is given by a Padovan sequence:
\begin{equation}
\sum_n x^n K_n=\frac{x^3}{1-x^2-x^3}\,;\quad
K_n=K_{n-2}+K_{n-3}\,;\quad K_1=0\,;\quad K_2=0\,;\quad K_3=1\,,\label{Kn}
\end{equation}
which is developed in Table~1. Note that the dimension of
the search space for a general MZV of weight $n$ is $K_{n+3}$~\cite{DZ},
which exceeds $K_n$ by a factor~\cite{AS} of $2.324717957$, as $n\to\infty$.

\subsection*{3. Knot-numbers from evaluations of Feynman diagrams}

The methods at our disposal~\cite{DK1,DK2,BKP} do not yet permit us to
predict, {\em a priori\/}, the transcendental knot-number assigned to
a positive knot by field-theory counterterms; instead we need
a concrete evaluation of at least one diagram
which skeins to produce that knot.
Neither do they allow us to predict the rational coefficients with which
such knot-numbers, and their products, corresponding to factor knots,
occur in a counterterm; instead we must, at present, determine these
coefficients by analytical calculation, or by applying a lattice method,
such as  MPPSLQ~\cite{DHB}, to (very) high-precision numerical data.
Nonetheless, the consequences of~\cite{DK1,DK2} are highly predictive
and have survived intensive testing with amazing fidelity. The origin
of this predictive content is clear: once a knot-number is determined
by one diagram, it is then supposed, and indeed found, to occur in the
evaluation of all other diagrams which skein to produce that knot.
Moreover, the search
space for subdivergence-free counterterms that evaluate to MZVs
is impressively smaller than that for the MZVs themselves,
due to the absence of any knot associated with $\pi^2$,
and again the prediction is borne out by a wealth of data.

We exemplify these features by considering diagrams that
evaluate to MZVs of depths up to 5, which is the greatest depth
that can occur at weights up to 17, associated with knots up to
crossing-number 17, obtained from diagrams with up to 10 loops.
We follow the economical notation of~\cite{GPX,DJB,BKP}, referring
to a vacuum diagram by a so-called angular diagram~\cite{GPX}, which
results from choosing one vertex as origin, and indicating all vertices
that are connected to this origin by dots, after removing the origin
and the propagators connected to it.
{From} such an angular diagram one may uniquely reconstruct
the original Feynman diagram. The advantage of this notation is that
the Gegenbauer-polynomial $x$-space technique~\cite{GPX} ensures
that the maximum depth of sum which can result is the smallest
number of propagators in any angular diagram that characterizes
the Feynman diagram. Fig.~1 shows a naming convention for log-divergent
vacuum diagrams with angular diagrams that yield up to 5-fold sums.
To construct, for example, the 7-loop diagram $G(4,1,0)$ one places
four dots on line 1 and one dot on line 2.
Writing an origin at any point disjoint from the angular diagram,
and joining all 6 dots to that origin, one recovers the
Feynman diagram in question.
Using analytical techniques developed
in~\cite{GPX,DJB,BG,EUL}, we find that all subdivergence-free diagrams
of $G$-type, up to 13 loops (the highest number computable in
the time available), give counterterms that evaluate
to $\zeta(2n+1)$, their products, and depth-3 knot-numbers
chosen from the sets
\begin{eqnarray}
N_{2m+1,2n+1,2m+1}&=&
\zeta(2m+1,2n+1,2m+1)-\zeta(2m+1)\,\zeta(2m+1,2n+1)\nonumber\\&&{}
+\sum_{k=1}^{m-1}{2n+2k\choose2k}\zeta_P(2n+2k+1,2m-2k+1,2m+1)\nonumber\\&&{}
-\sum_{k=0}^{n-1}{2m+2k\choose2k}\zeta_P(2m+2k+1,2n-2k+1,2m+1)
\,,\label{K3o}\\
N_{2m,2n+1,2m}&=&
\zeta(2m,2n+1,2m)+\zeta(2m)\left\{\zeta(2m,2n+1)+\zeta(2m+2n+1)\right\}
\nonumber\\&&{}
+\sum_{k=1}^{m-1} {2n+2k\choose2k  }\zeta_P(2n+2k+1,2m-2k,2m)\nonumber\\&&{}
+\sum_{k=0}^{n-1} {2m+2k\choose2k+1}\zeta_P(2m+2k+1,2n-2k,2m)
\,,\label{K3e}
\end{eqnarray}
where $\zeta_P(a,b,c)=\zeta(a)\left\{2\,\zeta(b,c)+\zeta(b+c)\right\}$.
The evaluation of a 9-loop non-planar example, $G(3,2,2)$,
is given in~\cite{EUL}: it evaluates in terms of MZVs of weights
ranging from 6 to 14. Choosing from\Eqq{K3o}{K3e}
one knot-number at 11 crossings and two at 13 crossings,
one arrives at an expression from which all powers of $\pi^2$ are banished,
which is a vastly more specific result than for a generic collection of MZVs
of these levels, and is in striking accord with what is required by the
knot-to-number connection entailed by field theory. Moreover, all planar
diagrams that evaluate to MZVs have been found to contain terms purely of
weight $2L-3$ at $L$ loops, matching the pattern of the zeta-reducible
crossed ladder diagrams~\cite{DK1,DK2}.

Subdivergence-free counterterms obtained from the $M$-type angular diagrams of
Fig.~1 evaluate to MZVs of weight $2L-4$, at $L$-loops, with depths up to 4.
Up to $L=8$ loops, corresponding to 12 crossings, the depth-4 MZVs can be
reduced to the depth-2 alternating sums~\cite{EUL}
$N_{a,b}\equiv\zeta(\overline{a},b)-\zeta(\overline{b},a)$. The knot-numbers
for the $(4,3)$ and $(5,3)$ torus knots may be taken as $N_{5,3}$
and $N_{7,3}$, thereby banishing $\pi^2$ from the associated 6-loop
and 7-loop counterterms. In general, $N_{2k+5,3}$ is a $(2k+8)$-crossing
knot-number at $(k+6)$ loops. Taking the second knot-number at 12 crossings
as~\cite{BGK,EUL} $N_{7,5}-\frac{\pi^{12}}{2^5 10!}$,
we express all 8-loop $M$-type counterterms in a $\pi$-free form.
At 9 loops, and hence 14 crossings, we encounter the first depth-4 MZV
that cannot be pushed down to alternating Euler sums of depth 2.
The three knot-numbers are again highly specific: to $N_{11,3}$
we adjoin
\begin{equation}
N_{9,5}+\df{5\pi^{14}}{7032946176}\,;\quad
\zeta(5,3,3,3)+\zeta(3,5,3,3)-\zeta(5,3,3)\zeta(3)
+\df{24785168\pi^{14}}{4331237155245}\,.\label{k14}
\end{equation}
Having determined these knot-numbers by applying MPPSLQ to
200 significant-figure evaluations of two counterterms, in a search space
of dimension $K_{17}=21$, requisite for generic MZVs of weight 14,
knot theory requires that we find the remaining five $M$-type counterterms
in a search space of dimension merely $K_{14}=9$. This prediction is totally
successful. The rational coefficients are too cumbersome to write here;
the conclusion is clear: when counterterms evaluate to MZVs they live
in a $\pi$-free domain, much smaller than that inhabited by a generic
MZV, because of the apparently trifling circumstance that a knot with
only two crossings is necessarily the unknot.

Such wonders continue, with subdivergence-free diagrams of types
$C$ and $D$ in Fig.~1
Up to 7 loops we have obtained {\em all\/} of them in terms of the
established knot-numbers $\{\zeta(3),\zeta(5),\zeta(7),N_{5,3},\zeta(9),
N_{7,3},\zeta(11),N_{3,5,3}\}$,
associated in~\cite{BKP,BGK} with the positive knots
$\{(3,2),(5,2),(7,2),8_{19}=(4,3),(9,2),10_{124}=(5,3),(11,2),11_{353}
=\sigma_1^{}\sigma_2^{3}\sigma_3^{2}\sigma_1^{2}\sigma_2^{2}\sigma_3^{}\}$,
and products of those knot-numbers, associated with the corresponding
factor knots.
A non-planar $L$-loop diagram may have terms of different weights,
not exceeding $2L-4$.
Invariably, a planar $L$-loop diagram evaluates purely
at weight $2L-3$.
Hence we expect the one undetermined MZV knot-number at
15 crossings to appear in, for example, the planar 9-loop diagram
$C(1,0,4,0,1)$. To find the precise combination of
$\zeta(9,\overline3,\overline3)-\frac{3}{14}\zeta(7,\overline5,\overline3)$
with other Euler sums would require a search in a space of dimension
$K_{18}=28$. Experience suggests that would require an evaluation
of the diagram to about 800 sf, which is rather ambitious,
compared with the 200 sf which yielded\Eq{k14}. Once the number is found,
the search space for further counterterms shrinks to dimension $K_{15}=12$.

\subsection*{4. Positive knots associated with irreducible MZVs}

Table~2 gives the braid words~\cite{VJ} of 5 classes of positive knot.
For each type of knot, ${\cal K}$,
we used the skein relation to compute the HOMFLY polynomial~\cite{VJ},
$X_{\cal K}(q,\la)$, in terms of
$p_n=(1-q^{2n})/(1-q^2)$, $r_n=(1+q^{2n-1})/(1+q)$,
$\Lambda_n=\la^n(1-\la)(1-\la q^2)$.

\noindent{\bf Table~2}:
Knots and HOMFLY polynomials associated with irreducibles MZVs.
\[\begin{array}{|l|l|l|}\hline{\cal K}&X_{\cal K}(q,\la)\\\hline
{\cal T}_{2k+1}=\sigma_1^{2k+1}&T_{2k+1}=\la^k(1+q^2(1-\la)p_k)\\
{\cal R}_{k,m}=\sigma_1^{}\sigma_2^{2k+1}\sigma_1^{}\sigma_2^{2m+1}&
R_{k,m}= T_{2k+2m+3}+q^3p_k p_m\Lambda_{k+m+1}\\
{\cal R}_{k,m,n}=
\sigma_1^{}\sigma_2^{2k}\sigma_1^{}\sigma_2^{2m}\sigma_1^{}\sigma_2^{2n+1}&
R_{k,m,n}=R_{1,k+m+n-1}+q^6p_{k-1}p_{m-1}r_n\Lambda_{k+m+n+1}\\
{\cal S}_{k}=
\sigma_1^{}\sigma_2^{3}\sigma_3^{2}\sigma_1^{2}\sigma_2^{2k}\sigma_3^{}&
S_{k}= T_3^2T_{2k+3}+q^2p_k r_2(q^2(\la-2)+q-2)\Lambda_{k+3}\\
{\cal S}_{k,m,n}=
\sigma_1^{}\sigma_2^{2k+1}\sigma_3^{}\sigma_1^{2m}\sigma_2^{2n+1}\sigma_3^{}
&S_{k,m,n}=T_{2k+2m+2n+3}+q^3(p_k p_m+p_m p_n+p_n p_k\\&\phantom{S_{k,m,n}=}
\quad{}+(q^2(3-\la)-2q)p_k p_m p_n)\Lambda_{k+m+n+1}\\\hline
\end{array}\]

Noting that ${\cal S}_{1,1,1}={\cal S}_{1}$ and
${\cal S}_{m,n,0}={\cal R}_{m,n,0}={\cal R}_{m,n}$,
one easily enumerates the knots of Table~2. The result is given,
up to 17 crossings, in the last row of Table~3,
where it is compared with the enumeration of all prime knots, which is known
only to 13 crossings, and with the enumeration of positive knots,
which we have
achieved to 15 crossings, on the assumption that the HOMFLY polynomial
has no degeneracies among positive knots. It is apparent
that positive knots are sparse, though they exceed the irreducible
MZVs at 10 crossings and at all crossing numbers greater than 11.
The knots of Table 2 are equal in number to the irreducible MZVs up to
16 crossings; thereafter they are deficient.
Table~4 records a finding that may be new: the Alexander
polynomial~\cite{VJ}, obtained by setting $\la=1/q$ in the HOMFLY polynomial,
is not faithful for positive knots. The Jones polynomial~\cite{VJ}, with
$\la=q$, was not found to suffer from this defect.
Moreover, by using REDUCE~\cite{RED}, and assuming the fidelity of the
HOMFLY polynomial in the positive sector, we were able to prove,
by exhaustion, that none of the $4^{14}\approx2.7\times10^8$ positive
5-braid words of length 14 gives a true 5-braid 14-crossing knot.

\noindent{\bf Table~3}:
Enumerations of classes of knots by crossing number, $n$,
compared with\Eq{Ms}.
\[\begin{array}{|r|rrrrrrrrrrrrrrr|}\hline
n&3&4&5&6&7&8&9&10&11&12&13&14&15&16&17\\\hline
\mbox{prime knots}&1&1&2&3&7&21&49&165&552&2176&9988&?&?&?&?\\
\mbox{positive knots}&1&0&1&0&1&1&1&3&2&7&9&17&47&?&?\\
M_n&1&0&1&0&1&1&1&1&2&2&3&3&4&5&7\\
\mbox{Table~2 knots}&1&0&1&0&1&1&1&1&2&2&3&3&4&5&5\\\hline
\end{array}\]

\noindent{\bf Table~4}:
Pairs of positive knots with the same Alexander polynomial,
$X_{\cal K}(q,1/q)$.
\[\begin{array}{|l|l|l|}\hline{\cal K}_1&{\cal K}_2&
X_{{\cal K}_1}(q,\la)-X_{{\cal K}_2}(q,\la)\\\hline
{\cal S}_{2,1,2}=
\sigma_1^{}
\sigma_2^{5}
\sigma_3^{}
\sigma_1^{2}
\sigma_2^{5}
\sigma_3^{}& 
\sigma_1^{3}
\sigma_2^{4}
\sigma_3^{}
\sigma_1^{2}
\sigma_2^{2}
\sigma_3^{2}
\sigma_2^{}& 
q^4(1-\la q)p_2r_2\Lambda_6\\
(\sigma_1^{}
\sigma_2^{2}
\sigma_3^{})^2
\sigma_1^{}
\sigma_2^{5}
\sigma_3^{}
& 
(\sigma_1^{}
\sigma_2^{2}
\sigma_3^{})^2
\sigma_1^{3}
\sigma_2^{}
\sigma_1^{2}
\sigma_3^{}& 
q^5(1-\la q)p_2\Lambda_6\\
\sigma_1^{5}
\sigma_2^{}
\sigma_3^{}
\sigma_1^{2}
\sigma_2^{3}
\sigma_3^{2}
\sigma_2^{}& 
\sigma_1^{2}
\sigma_2^{2}
\sigma_1^{3}
\sigma_2^{7}& 
q^5(1-\la q)p_2\Lambda_6\\\hline
\end{array}\]

The association~\cite{DK1,DK2} of the 2-braid torus knots
$(2k+1,2)={\cal T}_{2k+1}$
with the transcendental numbers $\zeta(2k+1)$ lies at the heart of our work.
In~\cite{DK2,BKP}, we associated the 3-braid torus knot
$(4,3)=8_{19}={\cal R}_{1,1}$ with the unique irreducible MZV
at weight 8, and in~\cite{BKP} we associated $(5,3)=10_{124}={\cal R}_{2,1}$
with that at weight 10. The 7-loop counterterms of $\phi^4$-theory
indicate that the knot-numbers associated with
$10_{139}=\sigma_1^{}\sigma_2^{3}\sigma_1^{3}\sigma_2^{3}$
and
$10_{152}=\sigma_1^{2}\sigma_2^{2}\sigma_1^{3}\sigma_2^{3}$
are not~\cite{BKP} MZVs.
At 11 crossings, the association of the knot-number $N_{3,5,3}$
with  ${\cal S}_1={\cal S}_{1,1,1}\equiv11_{353}$ is rock solid:
we have obtained
this number analytically from 2 diagrams and numerically from another 8,
in each case finding it with different combinations of $\zeta(11)$
and the factor-knot transcendental $\zeta^2(3)\zeta(5)$.
In~\cite{BGK} we associated the family of knots ${\cal R}_{k,m}$
with the knot-numbers $N_{2k+3,2m+1}$, modulo multiples of $\pi^{2k+2m+4}$
that have now been determined up to 14 crossings.
It therefore remains to explain here how:
(a) two families of 4-braids, ${\cal S}_{k}$ and ${\cal S}_{k,m,n}$,
diverge from their common origin at 11 crossings, to give two knots
at 13 crossings, and three at 15 crossings, associated with triple Euler sums;
(b) a new family, ${\cal R}_{k,m,n}$, begins at 14 crossings, giving
the $(7,3)$ torus knot, $(\sigma_1^{}\sigma_2^{})^7
=(\sigma_1^{}\sigma_2^4)^2\sigma_1^{}\sigma_2^3
={\cal R}_{2,2,1}$, associated with a truly irreducible four-fold sum.

To relate the positive knots of Table 2 to Feynman diagrams
that evaluate to MZVs we shall
dress their braid words with chords. In each of Figs.~2--8, we
proceed in two stages: first we extract, from a braid word,
a reduced Gauss code that defines a trivalent chord diagram;
then we indicate how to shrink propagators to obtain a scalar diagram that
is free of subdivergences and has an overall divergence
that evaluates to MZVs. Our criterion for reducibility to MZVs is
that there be an angular diagram, obtained~\cite{GPX,DJB}
by choosing one vertex as an origin, such that the angular
integrations may be performed without encountering
6--j symbols, since these appeared in all the diagrams involving the
non-MZV knots $10_{139}$ and $10_{152}$ at 7 loops~\cite{BKP}, whereas
all the MZV-reducible diagrams could be cast in a form free of 6--j
symbols.

The first step -- associating a chord diagram with a knot --
allows considerable freedom: each
chord is associated with a horizontal propagator connecting vertical
strands of the braid between crossings, and there are almost
twice as many crossings as there are chords in the corresponding
diagram. Moreover, there are several braid words representing the same knot.
Thus a knot can be associated with several chord diagrams. Figs.~3b and~4b
provide an example: each diagram obtained from the $(5,2)$ torus knot
yields a counterterm involving $\zeta(5)$,
in a trivalent theory such as QED or Yukawa theory.

In Table 2 we have five families of braid words: the 2-braid torus knots,
two families of 3-braids, and two families of 4-braids. We begin with
the easiest family, ${\cal T}_{2k+1}$.
Consider Fig.~2a. We see the braid $\sigma_1^3$, dressed with two
horizontal propagators. Such propagators will be called chords,
and we shall refer to Figs.~2a--8a as chorded braid diagrams.
In Fig.~2a the two chords are labelled 1 and 2.
Following the closed braid, starting from the upper end
of the left strand, we encounter each chord twice, at vertices
which we label $\{1,{1^\prime}\}$ and $\{2,{2^\prime}\}$. These
occur in the order
$1,2,{1^\prime},{2^\prime}$ in Fig.~2a. This is the same order as they
are encountered on traversing the circle of Fig.~2b, which
is hence the same diagram as the chorded braid of Fig.~2a.
As a Feynman diagram, Fig.~2b is indeed associated with
the trefoil knot, by the methods of~\cite{DK1}.
We shall refer to the Feynman diagrams of Figs.~2b--8b as
chord diagrams\footnote{The reader familiar with
recent developments in knot theory and topological field theory might
notice that our notation is somewhat motivated by the connection between
Kontsevich integrals~\cite{LM} and chord diagrams. In~\cite{DK}
this will be discussed in detail and related to the work in~\cite{DK1}.}.
Each chord diagram is merely
a rewriting of the chorded braid diagram that precedes it,
displaying the vertices
on a hamiltonian circuit that passes through all the vertices.
The final step is trivial in this example:
the scalar tetrahedron is already log-divergent in 4 dimensions, so no
shrinking of propagators is necessary. Fig.~2c records the trivial
angular diagram, obtained~\cite{GPX} by choosing ${2}$ as an origin
and removing the propagators connected to it:
this merely represents a wheel with 3 spokes. In general~\cite{DJB}
the wheel with $n+1$ spokes delivers $\zeta(2n-1)$.

In Fig.~3a we give a chording of the braid $\sigma_1^{2n-1}$,
which is the simplest representation of the $(2n-1,2)$ torus knot,
known from previous work~\cite{DK1,DK2} to be associated with
a $(n+1)$-loop diagram, and hence with a hamiltonian circuit that has $n$
chords. Thus each addition of $\sigma_1^2$ involves adding a
single chord, yielding the chord diagram of Fig.~3b. To obtain
a logarithmically divergent scalar diagram, we shrink the propagators
connecting vertex ${2^\prime}$ to vertex ${n^\prime}$, drawn with a thick
line on the hamiltonian circuit of Fig.~3b, and hence obtain
the wheel with $n+1$ spokes, represented by the angular diagram of Fig.~3c.

To show how different braid-word representations of the
same knot give different chord diagrams, yet the same transcendental,
we consider Fig.~4.
In Fig.~4a we again have a chorded braid with $n$ chords, which this time
is obtained by combining $\sigma_1\sigma_2\sigma_1\sigma_2$
with $n-2$ powers of $\sigma_2^2$.
The resultant braid word,
$\sigma_1^{}\sigma_2^{}\sigma_1^{}\sigma_2^{2n-3}$,
is the $(2n-1,2)$ torus knot written as a 3-braid.
Labelling the pairs of vertices of Fig.~4b, one sees that it is identical
to the closure of the braid of Fig.~4a.
Shrinking together the vertices
$\{{2^\prime},{n^\prime},\ldots,{3^\prime}\}$
gives the angular diagram of Fig.~4c, which is
the same as Fig.~3c and hence delivers $\zeta(2n-1)$.

This ends our consideration of the 2-braid torus knots. We now turn
to the class ${\cal R}_{k,m}$ in Fig.~5.
The first member ${\cal R}_{1,1}=8_{19}=(4,3)$ appears at 6 loops,
with five chords.
It was obtained from Feynman diagrams in \cite{DK2}, and found
in~\cite{BKP} to be associated with an MZV of depth 2.
In Fig.~5a we add singly-chorded powers of $\sigma_2^2$
to a chorded braid word that delivers a Feynman diagram for which the
procedures of~\cite{DK1} gave $8_{19}$ as one of its skeinings.
In general, we have $k+m+3$ chords and thus $k+m+4$ loops.
The resulting chord diagram
is Fig.~5b, whose 7-loop case was the basis for associating
$10_{124}$ with the MZV $\zeta(7,3)$~\cite{BKP}.
Shrinking the propagators indicated by thickened lines in Fig.~5b,
we obtain diagram $M(k,1,1,m)$, indicated by the angular diagram of
Fig.~5c. Explicit computation of all such diagrams, to 9
loops, shows that this family is indeed MZV-reducible, to 14 crossings.

In Fig.~6 we repeat the process of Fig.~5 for the knot class
${\cal R}_{k,m,n}$. Marked boxes, in Fig.~6a,
indicate where we increase the number of chords.
Fig.~6b shows the highly non-planar chord diagram
for this knot class. This non-planarity is maintained
in the log-divergent diagram obtained by shrinking the thickened lines
in Fig.~6b. The parameters $m$ and $k$ correspond to the
 series of dots in the corresponding angular diagram of Fig.~6c.
Non-planarity is guaranteed by the two remaining dots,
which are always present.
For $n>1$, we see even more propagators in the angular diagram.
The absence of 6--j symbols from angular integrations leads us to believe
that the results are reducible to MZVs; the non-planarity entails
MZVs of even weight, according to experience up to 7 loops~\cite{BKP}.

We now turn to the last two classes of knots: the 4-braids of Table~2.
In Fig.~7a we give a chorded braid diagram for knots
of class ${\cal S}_k$. Again, the marked box indicates
how we add chords to a chorded braid diagram
that corresponds to a 7-loop Feynman diagram, already known~\cite{BKP}
to skein to produce ${\cal S}_1=11_{353}$.
Shrinking the thickened lines in Fig.~7b, we obtain a log-divergent
planar diagram containing: a six-point coupling,
a $(k+3)$-point coupling, and $k+5$ trivalent vertices.
This is depicted in Fig.~7c as an angular diagram obtained by
choosing the $(k+3)$-point coupling as an origin.
Choosing the 6-point coupling as an origin for
the case $k=1$ confirms that ${\cal S}_1=11_{353}$ is associated
with $\zeta(3,5,3)$ via the 7-loop diagram $G(4,1,0)$.
However, for $k=3$ there is no way of obtaining MZVs of depth
3 from either choice of 6-point origin. Hence we expect a depth-5
MZV to be associated with the 15-crossing knot ${\cal S}_3$, with the
possibility of depth-7 MZVs appearing at higher crossings.

Finally we show that the three-parameter class
${\cal S}_{k,m,n}$, also built on $11_{353}={\cal S}_{1,1,1}$,
is associated with depth-3 MZVs.
The chorded braid of Fig.~8a indicates
the three places where we can add further chords.
Fig.~8b gives the chord diagram associated with it,
and indicates how to shrink propagators to obtain a log-divergent
diagram, represented by the angular diagram $G(m+n+2,k,0)$
of Fig.~8c, which evaluates in terms of depth-3 MZVs
up to 13 loops, and presumably beyond.

\subsection*{5. Conclusions}

In summary, we have
\begin{enumerate}
\item enumerated in\Eqq{Es}{Ms} the irreducibles entailed by Euler sums
and multiple zeta values at weight $n$; apportioned them by depth in\Eq{EMb}; 
conjectured the generator\Eq{Pd} for the number, $D_{n,k}$, of MZVs of weight 
$n$ and depth $k$ that are irreducible to MZVs of lesser depth; 
\item determined all MZV knot-numbers to 15 crossings, save one,
associated with a 9-loop diagram that evaluates to MZVs of depth 5
and weight 15;
\item enumerated positive knots to 15 crossings, notwithstanding
degenerate Alexander polynomials at 14 and 15 crossings;
\item developed
a technique of chording braids so as to generate families of knots
founded by parent knots whose relationship to Feynman diagrams
was known at lower loop numbers;
\item combined all the above to identify,
in Table~2, knots whose enumeration, to 16 crossings, matches that of MZVs.
\end{enumerate}
Much remains to be clarified in this rapidly developing area.
Positive knots, and hence the transcendentals associated with them
by field theory, are richer in structure than MZVs: there are more of them
than MZVs; yet those whose knot-numbers are MZVs evaluate in search
spaces that are significantly smaller than those for the MZVs, due to
the absence of a two-crossing knot. After 18 months of intense collaboration,
entailing large scale computations in knot theory, number theory, and field
theory, we are probably close to the boundary of what can be discovered
by semi-empirical methods. The trawl, to date, is impressive, to our minds.
We hope that colleagues will help us to understand it better.

\noindent{\bf Acknowledgements}
We are most grateful to Don Zagier for his generous comments,
which encouraged us to believe in the correctness of our discoveries\Eq{EMb},
while counselling caution as to the validity of\Eq{Pd}
in so far uncharted territory with depth $k>4$.
David Bailey's MPPSLQ~\cite{DHB}, Tony Hearn's REDUCE~\cite{RED}
and Neil Sloane's superseeker~\cite{NJAS} were instrumental in this work.
We thank Bob Delbourgo for his constant encouragement.

\newpage

\setlength{\unitlength}{0.014cm}
\newbox\shell
\newcommand{\lbl}[3]{\put(#1,#2){\makebox(0,0)[b]{$#3$}}}
\newcommand{\dia}[1]{\setbox\shell=\hbox{\begin{picture}(180,200)(-90,-110)#1
\end{picture}}\dimen0=\ht
\shell\multiply\dimen0by7\divide\dimen0by16\raise-\dimen0\box\shell}
\newcommand{\vtx}{\circle*{10}}

\noindent
Fig.~1: Angular diagrams yielding MZVs of depths up to 5.\\
\dia{
\put(0,0){\circle{100}}
\put(0,-50){\line(0,1){100}}
\put(0,50){\vtx}
\lbl{-60}{-5}{1}
\lbl{-10}{-5}{2}
\lbl{60}{-5}{3}
\lbl{0}{-90}{G(n_1,n_2,n_3)}
}\hfill
\dia{
\put(0,0){\oval(125,100)[r]}
\put(0,0){\oval(125,100)[l]}
\put(0,0){\oval(75,98)[r]}
\put(0,0){\oval(75,98)[l]}
\lbl{-72}{-5}{1}
\lbl{-25}{-5}{2}
\lbl{25}{-5}{3}
\lbl{72}{-5}{4}
\lbl{0}{-90}{M(n_1,\ldots,n_4)}
}\hfill
\dia{
\put(0,0){\circle{100}}
\put(0,50){\line(1,-2){40}}
\put(0,50){\line(-1,-2){40}}
\put(0,50){\vtx}
\lbl{-60}{-5}{1}
\lbl{-15}{-15}{2}
\lbl{0}{-45}{3}
\lbl{15}{-15}{4}
\lbl{60}{-5}{5}
\lbl{0}{-90}{C(n_1,\ldots,n_5)}
}\hfill
\dia{
\put(0,0){\circle{100}}
\put(0,50){\line(1,-2){40}}
\put(0,50){\line(-1,-2){40}}
\put(-40,-30){\vtx}
\lbl{-60}{-5}{1}
\lbl{-15}{-15}{2}
\lbl{0}{-45}{3}
\lbl{15}{-15}{4}
\lbl{60}{-5}{5}
\lbl{0}{-90}{D(n_1,\ldots,n_5)}
}\\[2cm]
Fig.~2: The lowest level non-trivial chorded braid appears at three
loops, and is given in (a). It is the same Feynman diagram as
the chord diagram in (b), and gives
rise to the simple angular diagram of (c).
In (a) we indicate the closure of the braid by dotted lines, which
we omit from subsequent figures.\nopagebreak\\
\epsfxsize=9cm\epsfbox{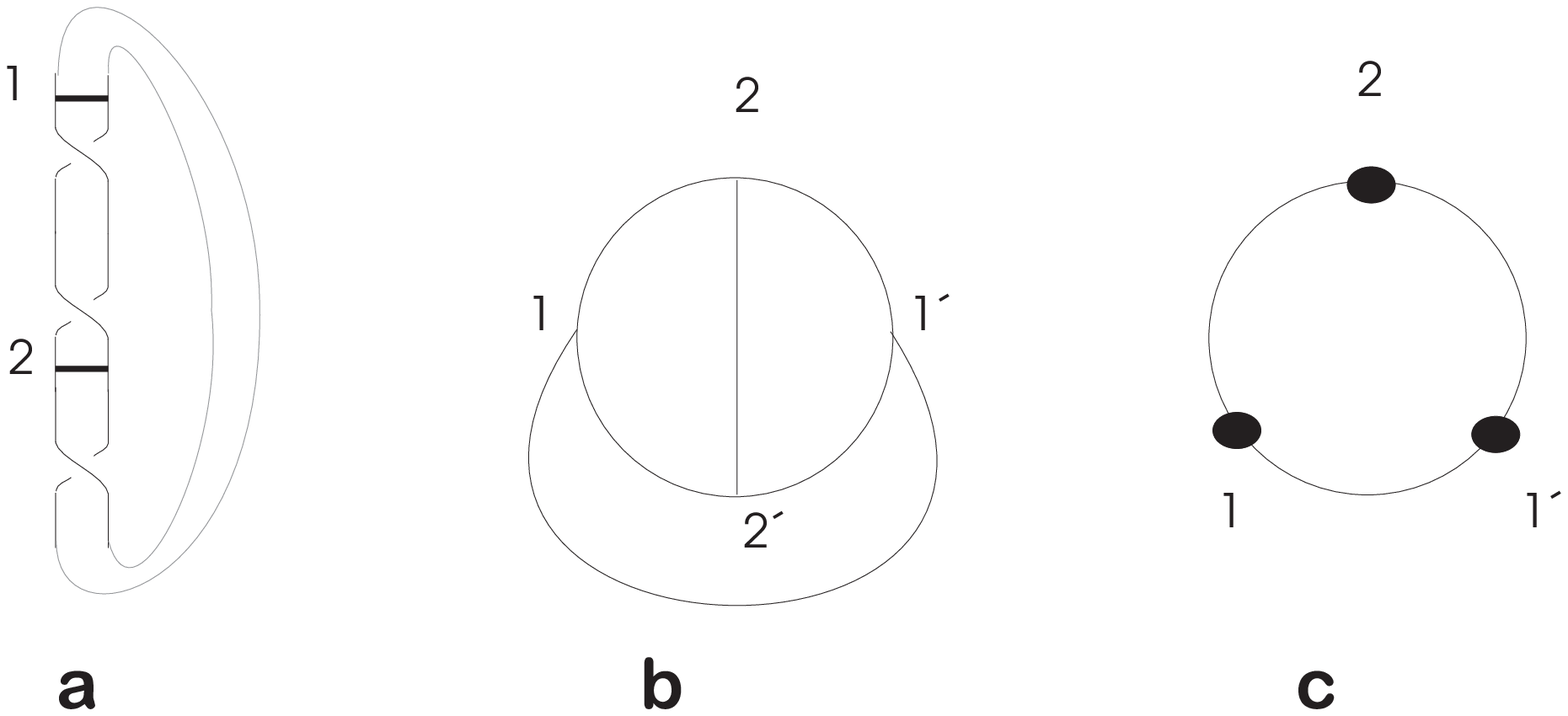}\\[2cm]
Fig.~3: We extend Fig.~2 to produce the chorded braid (a) and the
chord diagram (b),
which deliver $\zeta(2n-1)$ from the angular diagram of (c).
The thick line is shrunk to a point to produce a
$(n+1)$-point vertex, serving as the origin in (c).\nopagebreak\\[5mm]
\epsfxsize=9cm\epsfbox{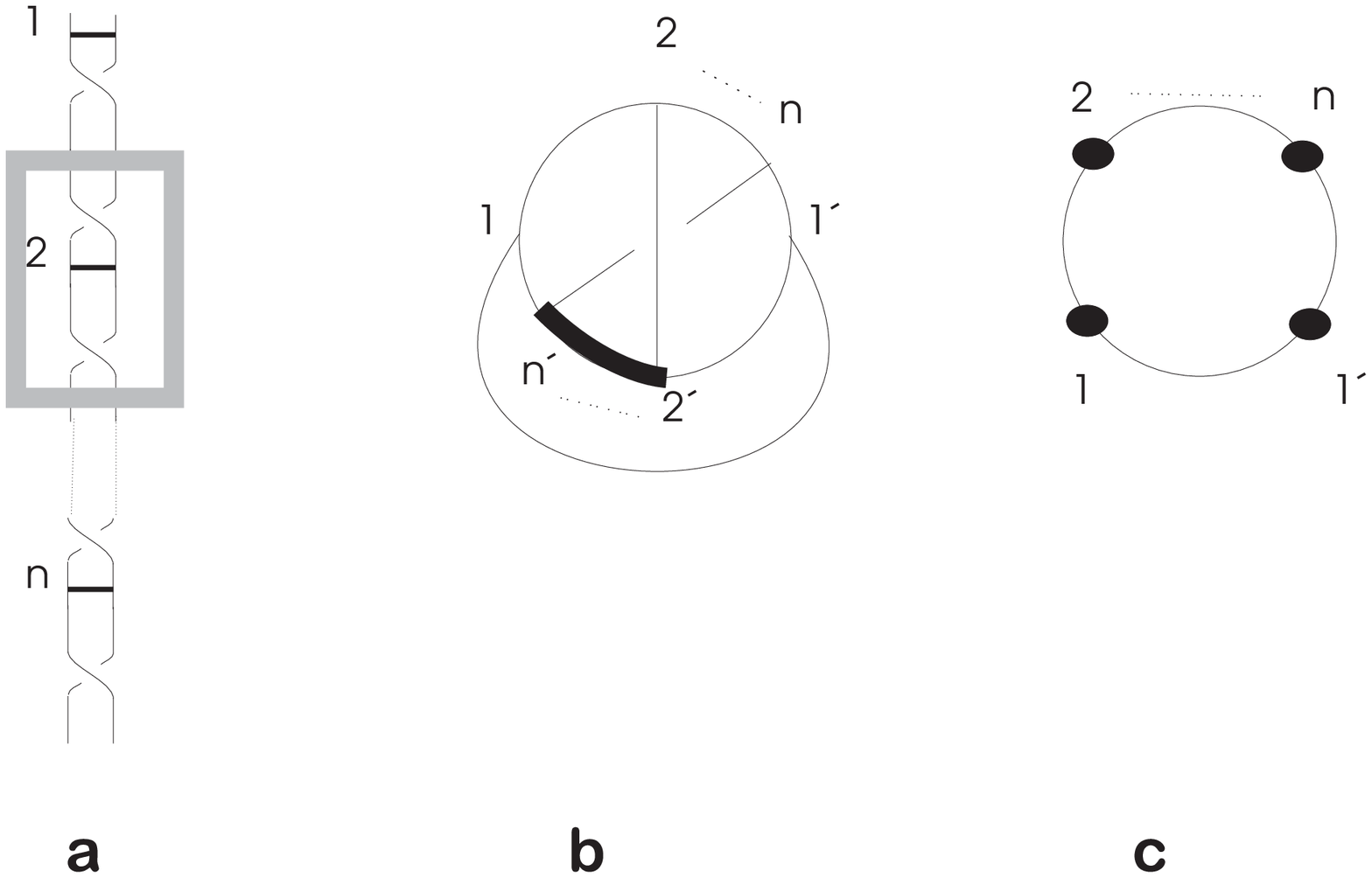}
\newpage
\noindent
Fig.~4: The braid in (a) is a 3-braid representation for the
$(2,2n-1)$ torus knot.
It produces an angular diagram (c) that is the same as in Fig.~3,
after shrinking the thickened line in (b).\nopagebreak\\[5mm]
\epsfxsize=9cm\epsfbox{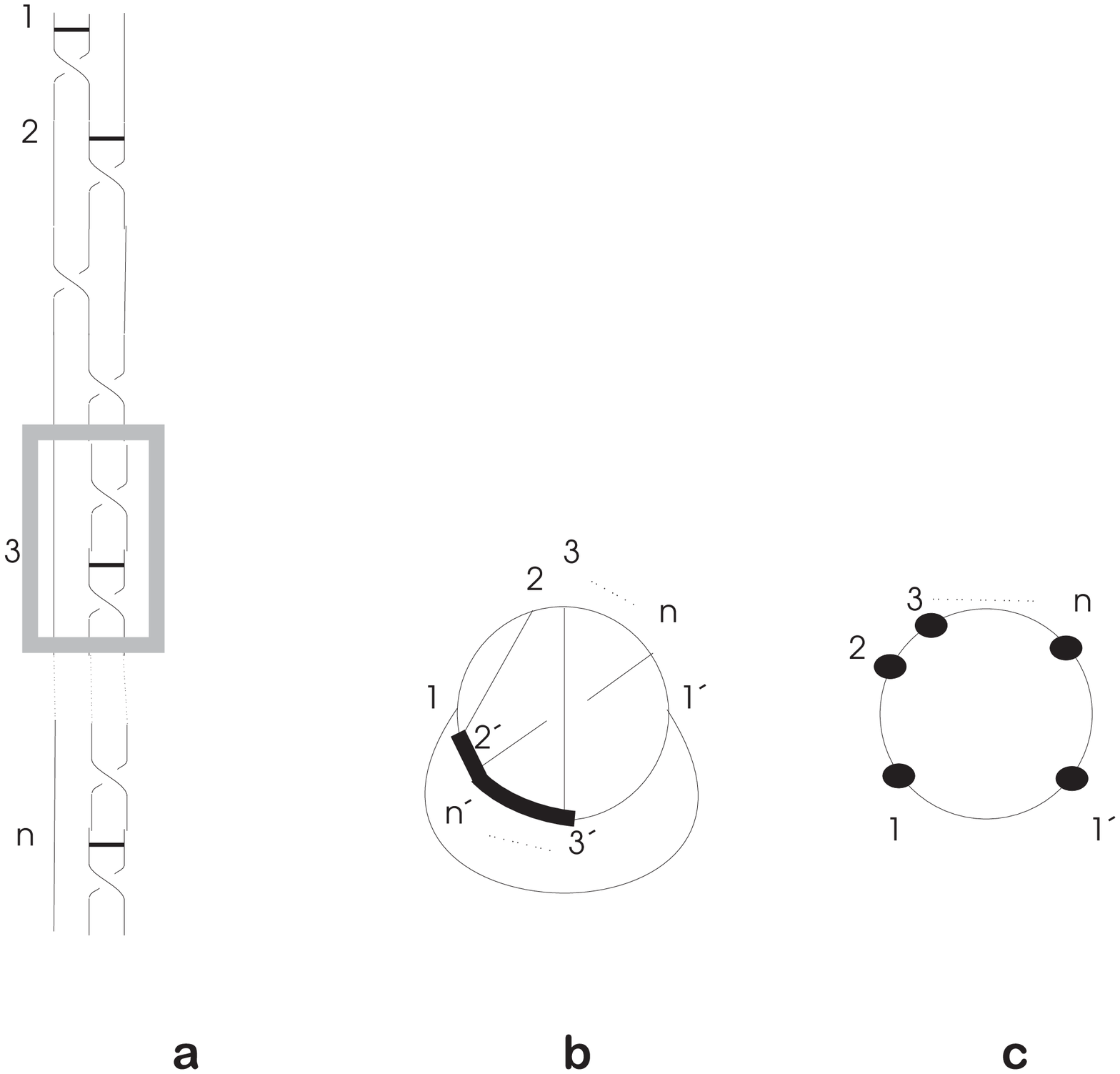}\\[1cm]
Fig.~5: In this genuine 3-braid example, the chorded braid
(a) is equivalent, as a Feynman diagram, to the chord diagram (b),
and gives rise to the class $M(k,1,1,m)$ of angular diagrams
in (c), by shrinking: two vertices to a 4-point vertex
at the top of the diagram in (b);
the propagator $(1,1^\prime)$ to a 4-point vertex;
and $k+m$ vertices to a $(k+m+2)$-point
vertex, which then serves as the origin in (c).\nopagebreak\\[5mm]
\epsfxsize=9cm\epsfbox{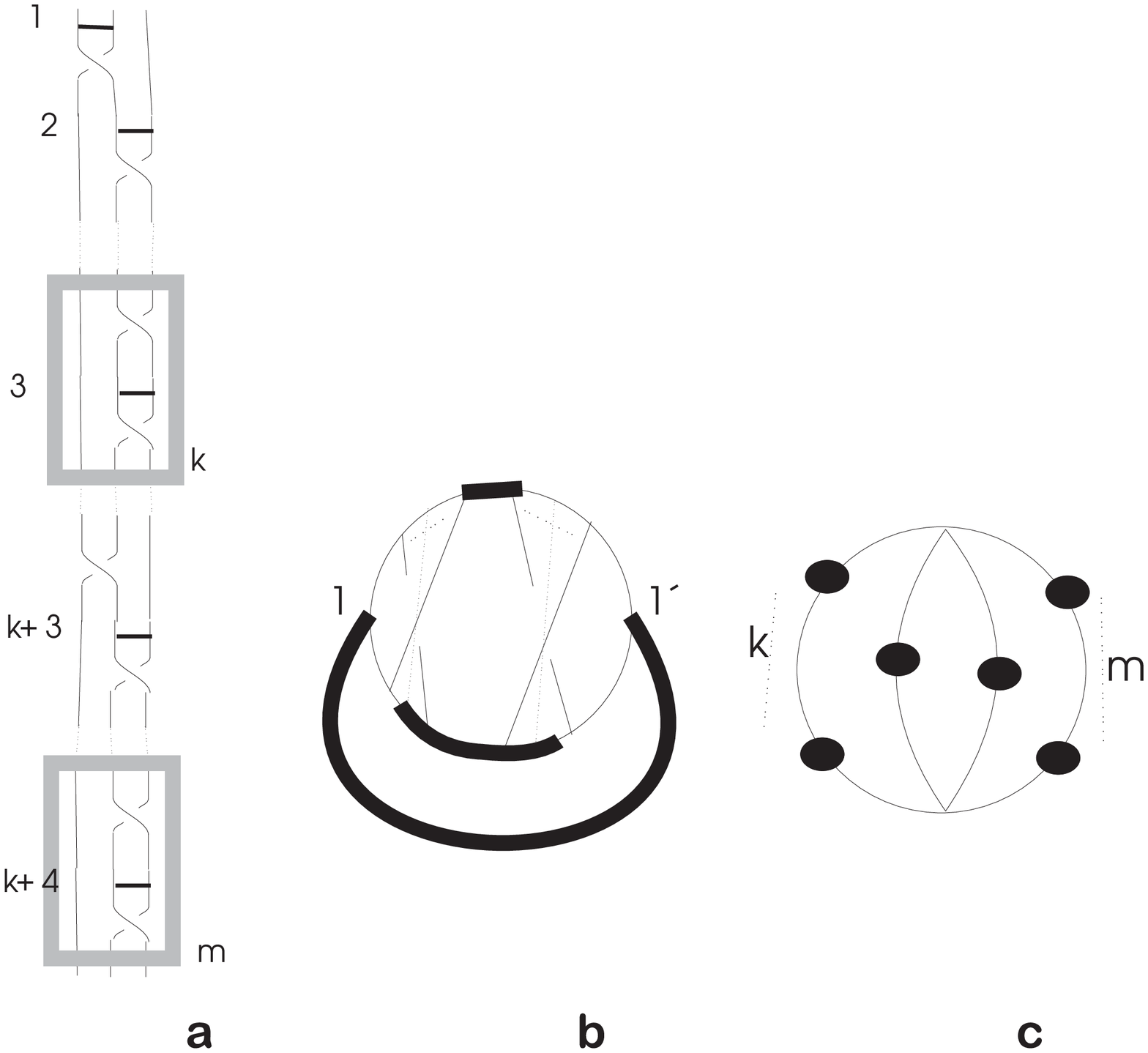}
\newpage
\noindent
Fig.~6: Another genuine class of 3-braids, which we believe
to entail MZVs of even weight.
We obtain two 4-point vertices by shrinking
the indicated thick lines at the top and right of the circle in (b).
We further shrink $k+m$ vertices at the bottom,
and $n+1$ vertices at the left, to obtain an angular
diagram (c) that is free of 6--j symbols.\nopagebreak\\[5mm]
\epsfxsize=9cm\epsfbox{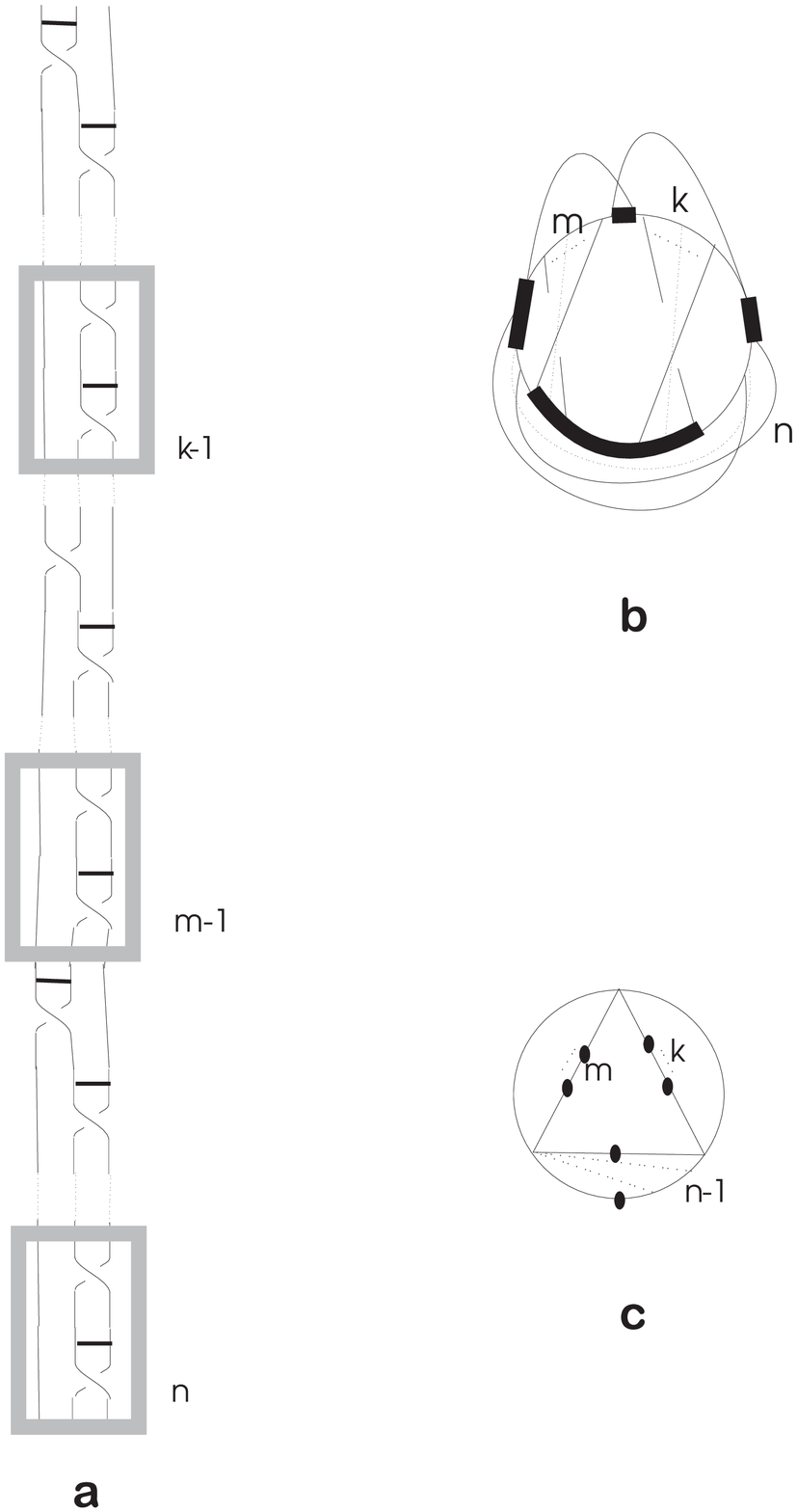}
\newpage
\noindent
Fig.~7: This shows the first class of genuine 4-braids,
which gives rise to a planar
angular diagram, beginning with $11_{353}$, for $k=1$.
We shrink four vertices at the top, and $k+1$
vertices at the right, as indicated in (b).\nopagebreak\\[5mm]
\epsfxsize=9cm\epsfbox{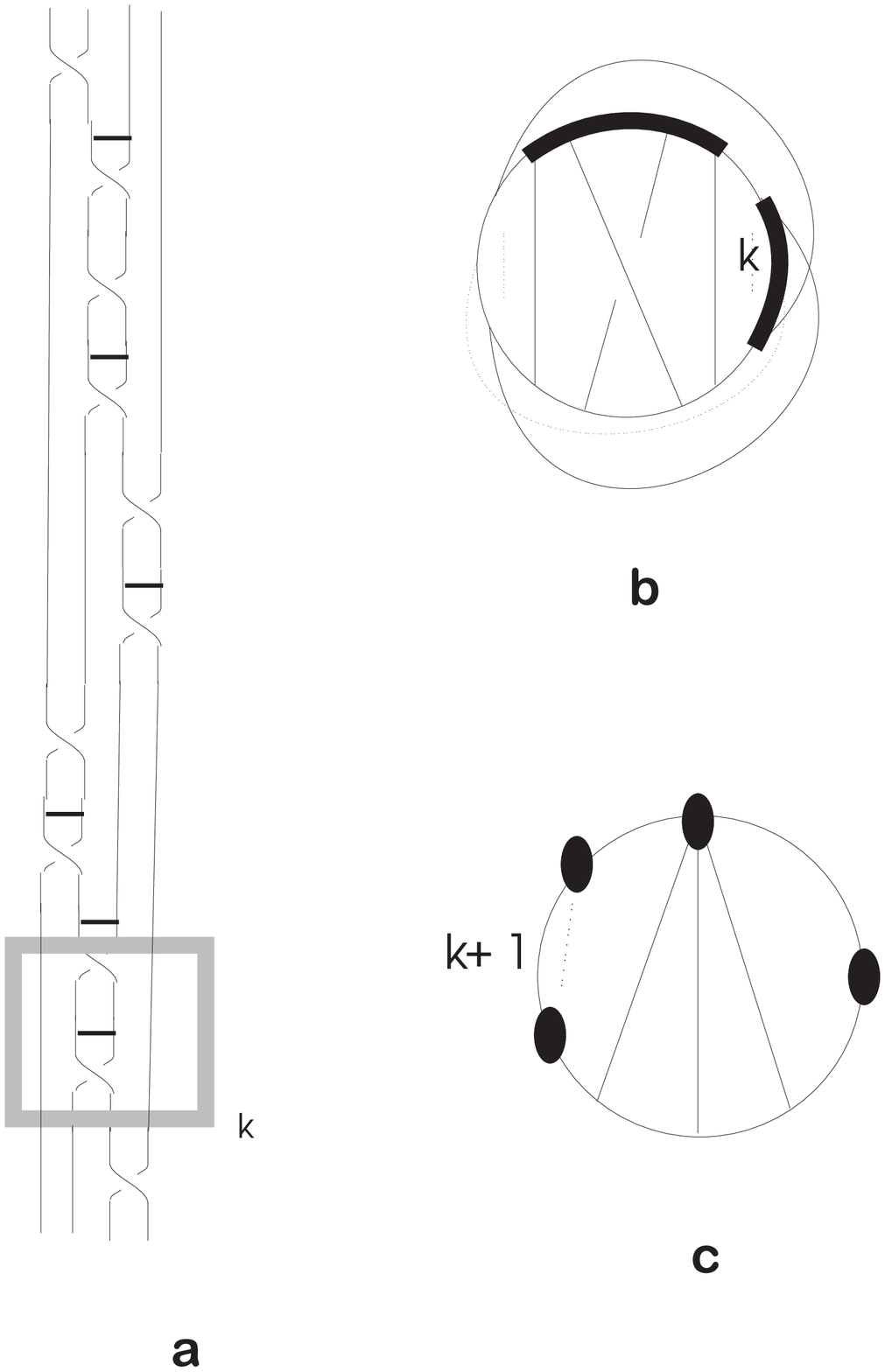}
\newpage
\noindent
Fig.~8: This triple-index class of 4-braids also begins with
$11_{353}$, for $k=m=n=1$.
{From} shrinking, we obtain a 4-point vertex at the left
of (b), and at the bottom a $(m+n+k+3)$-point vertex
that is the origin of the planar angular diagram $G(m+n+2,k,0)$ of (c).
\nopagebreak\\[5mm]
\epsfxsize=9cm\epsfbox{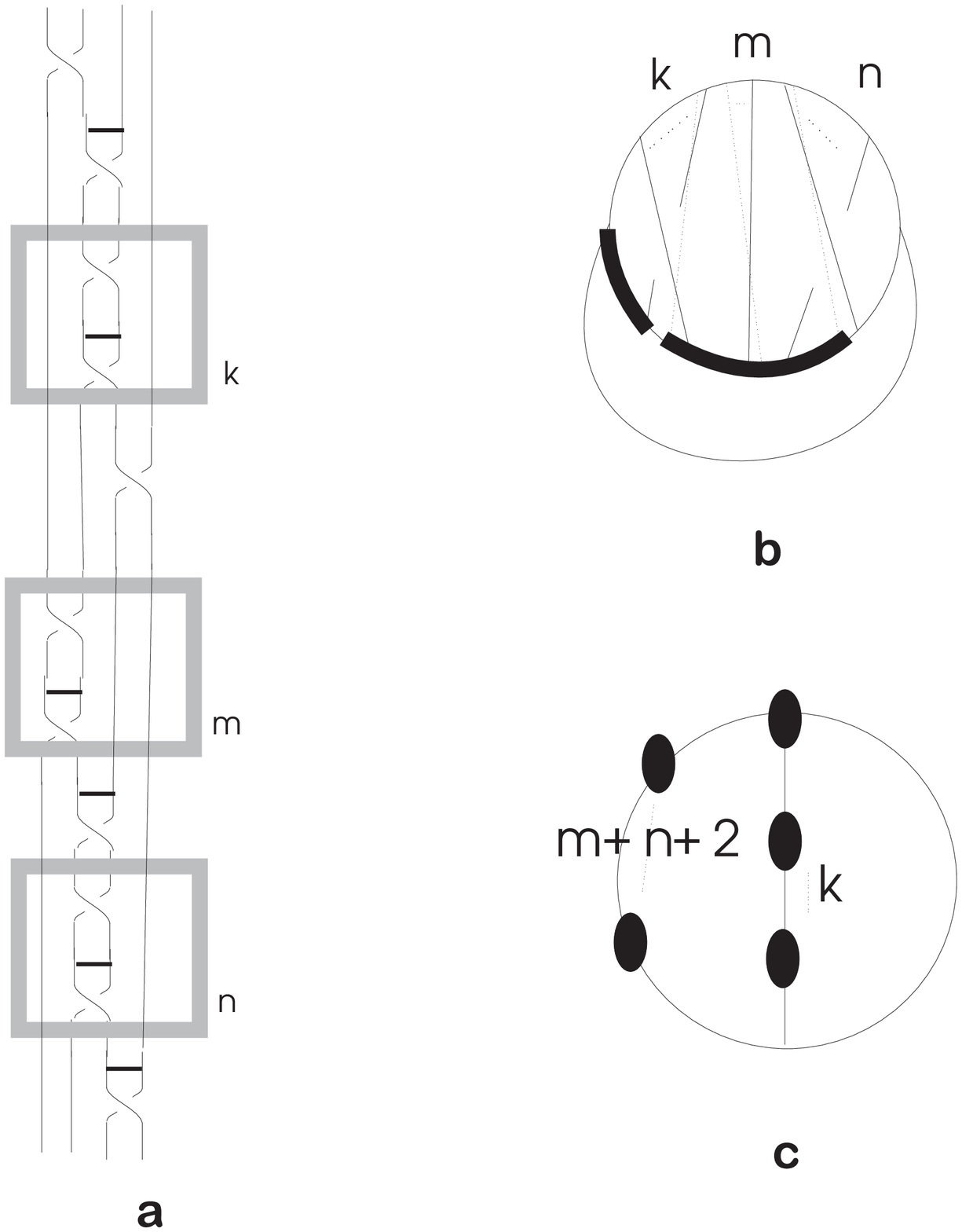}

\newpage
\begin{thebibliography}{99}

\bibitem{DK1}
D.\ Kreimer,
Habilitationsschrift: {\sl Renormalization and Knot Theory},
Mainz preprint MZ--TH--96--18 (July 1996),
to appear in Journal of Knot Theory and its Ramifications; q-alg/9607022.

\bibitem{DK2}
D.\ Kreimer,
Phys.\ Lett. {\bf B354} (1995) 117.

\bibitem{GPX}
K.G.\ Chetyrkin, A.L.\ Kataev and F.V.\ Tkachov,
Nucl.\ Phys. {\bf B174} (1980) 345;\\
H.\ Kleinert, J.\ Neu, V.\ Schulte-Frohlinde, K.G.\ Chetyrkin and
S.A.\ Larin,\\
Phys.\ Lett. {\bf B272} (1991) 39; {\bf B319} (1993) 545 (erratum).

\bibitem{DJB}
D.J.\ Broadhurst,
Open University report OUT-4102-18 (1985);
Phys.\ Lett. {\bf B164} (1985) 356;
Z.\ Phys. {\bf C32} (1986) 249;
Phys.\ Lett. {\bf B307} (1993) 132;\\
D.T.\ Barfoot and D.J.\ Broadhurst,
Z.\ Phys. {\bf C41} (1988) 81.

\bibitem{BKP}
D.J.\ Broadhurst and D.\ Kreimer,
Int.\ J.\ Mod.\ Phys. {\bf C6} (1995) 519.

\bibitem{DZ} 
D.\ Zagier, in Proc.\ First European Congress Math.\ (Birkh\"auser,
Boston, 1994) Vol II, pp 497-512; {\sl Multiple Zeta Values}, in preparation.

\bibitem{LM}
T.Q.T.\ Le and J.\ Murakami, MPI Bonn preprints 93-26; 93-89 (1993);\\
C.\ Kassel, {\sl Quantum groups} (Springer, New York, 1995) pp 480-483.

\bibitem{BDK}
D.J.\ Broadhurst, R.\ Delbourgo and D.\ Kreimer,
Phys.\ Lett. {\bf B366} (1996) 421.

\bibitem{BGK} 
D.J.\ Broadhurst, J.A.\ Gracey and D.\ Kreimer,
Open University preprint, OUT--4102--46 (1996); hep-th/9607174,
to appear in Z.\ Phys.\ C.

\bibitem{DKT} 
R.\ Delbourgo, A.\ Kalloniatis and G.\ P.\ Thompson,
Phys.\ Rev. {\bf D54} (1996) 5373.

\bibitem{BBG}
D.\ Borwein, J.M.\ Borwein and R.\ Girgensohn,
Proc.\ Edin.\ Math.\ Soc. {\bf 38} (1995) 273.

\bibitem{BG}
J.\ M.\ Borwein and R.\ Girgensohn,
Electronic J.\ Combinatorics {\bf 3} (1996) R23,
with an appendix by D.\ J.\ Broadhurst.

\bibitem{EUL}
D.J.\ Broadhurst,
Open University preprint, OUT--4102--62 (1996); hep-th/9604128, to appear
in J.\ Math.\ Phys.

\bibitem{BBB}  
J.M.\ Borwein, D.A.\ Bradley and D.J.\ Broadhurst,
Open University preprint, OUT--4102--63 (1996); hep-th/9611004,
to appear in Electronic J. Combinatorics.

\bibitem{DK}
D.\ Kreimer,
{\sl Knots and Feynman Diagrams}
(Cambridge University Press, in preparation).

\bibitem{VJ}
V.F.R.\ Jones,
Annals of Math. {\bf 126} (1987) 335.

\bibitem{AS}
W.\ Adams and D.\ Shanks,
Math.\ Comp. {\bf 39} (1982) 255.

\bibitem{DHB}
D.\ H.\ Bailey,
ACM Trans.\ Math.\ Software {\bf 21} (1995) 379.

\bibitem{RED}
A.C.\ Hearn,
REDUCE user's manual, version 3.5, Rand publication CP78 (1993).

\bibitem{NJAS}
N.\ J.\ A.\ Sloane,
Electronic J.\ Combinatorics {\bf 1} (1994) F1.

\end{thebibliography}
\end{document}